\documentclass[9pt,twocolumn,twoside]{opticajnl}
\journal{opticajournal} 

\setboolean{shortarticle}{true}

\usepackage{amsmath}
\usepackage{braket}
\usepackage{graphicx}
\usepackage{bm}
\usepackage{verbatim}
\usepackage{soul}
\usepackage{ulem}
\usepackage{xcolor}
\usepackage{xfrac}
\usepackage{lscape}

\newcommand*{\E}{\mathcal{E}}

\newcommand*{\MS}{\mathcal{S}}

\title{Polarization-entangled photon pair source using beam displacers and thin crystals}

\author[1,2,$\dagger$]{Minjae Hong}
\author[1,2]{Rodrigo G\'omez}
\author[1]{Valerio Flavio Gili}
\author[3]{Jorge Fuenzalida}
\author[1,3,*]{Markus Gräfe}

\affil[1]{Fraunhofer Institute for Applied Optics and Precision Engineering IOF, Albert-Einstein-Str. 7, 07745 Jena, Germany}
\affil[2]{Friedrich Schiller University Jena, Abbe Center of Photonics, Albert-Einstein-Str. 6, 07745 Jena, Germany}
\affil[3]{Institute for Applied Physics, Technical University of Darmstadt, Schloßgartenstraße 7, 64289 Darmstadt, Germany}

\affil[$\dagger$]{Present address: Max Planck Institute for the Science of Light, Staudtstraße 2; and Present address: Friedrich Alexander University, Schloßplatz 4, 91058 Erlangen, Germany}

\affil[*]{markus.graefe@tu-darmstadt.de}

\begin{abstract}
We present an experimental implementation of a polarization-entangled photon pair source based on beam displacers. The down-converted photons are emitted via spontaneous parametric down-conversion in a non-degenerate and type-0 process. We obtain a state fidelity of $F=0.975\pm0.004$ and violate a Clauser-Horne-Shimony-Holt inequality with $\MS=2.75\pm0.01$. Our source also uses thin crystals for applications in quantum imaging, taking advantage of the large number of spatial modes. We estimate that our source could produce 550$\pm$12 spatial modes.
\end{abstract}

\setboolean{displaycopyright}{false} 

\begin{document}

\maketitle

The generation of entangled photon pairs is of paramount importance for quantum technologies such as quantum imaging and sensing \cite{gilaberte2019perspectives,fuenzalida2024nonlinear} and quantum cryptography \cite{pirandola2020advances}. Spontaneous parametric down-conversion (SPDC) is the most well-known process for this task wherein a pump photon is down-converted to two photons, namely, signal and idler~\cite{klyshko1969scattering}. Down-converted photons offer entanglement in various degrees of freedom~\cite{walborn2010spatial} or entanglement can be prepared in a specific degree of freedom through the use of interferometers~\cite{anwar2021entangled}. By entangling more than one degree of freedom, the so-called hyper-entangled states are obtained. These states have several applications in different areas of quantum physics, ranging from distillation~\cite{ecker2021experimental}, and frequency~\cite{brambila2023ultrabright} and spatial multiplexing~\cite{ortega2021multicore,ortega2024implementation} in quantum communications to holography~\cite{defienne2021polarization} and super-sensitivity~\cite{camphausen2021quantum} in quantum imaging. 
Thus, these states deliver several advantages that, in most cases, cannot be achieved with only two qubits-entangled states. This work is a step towards a hyper-entangled state in polarization and space.

\begin{figure}[htpb]
\centering\includegraphics[width=8cm]{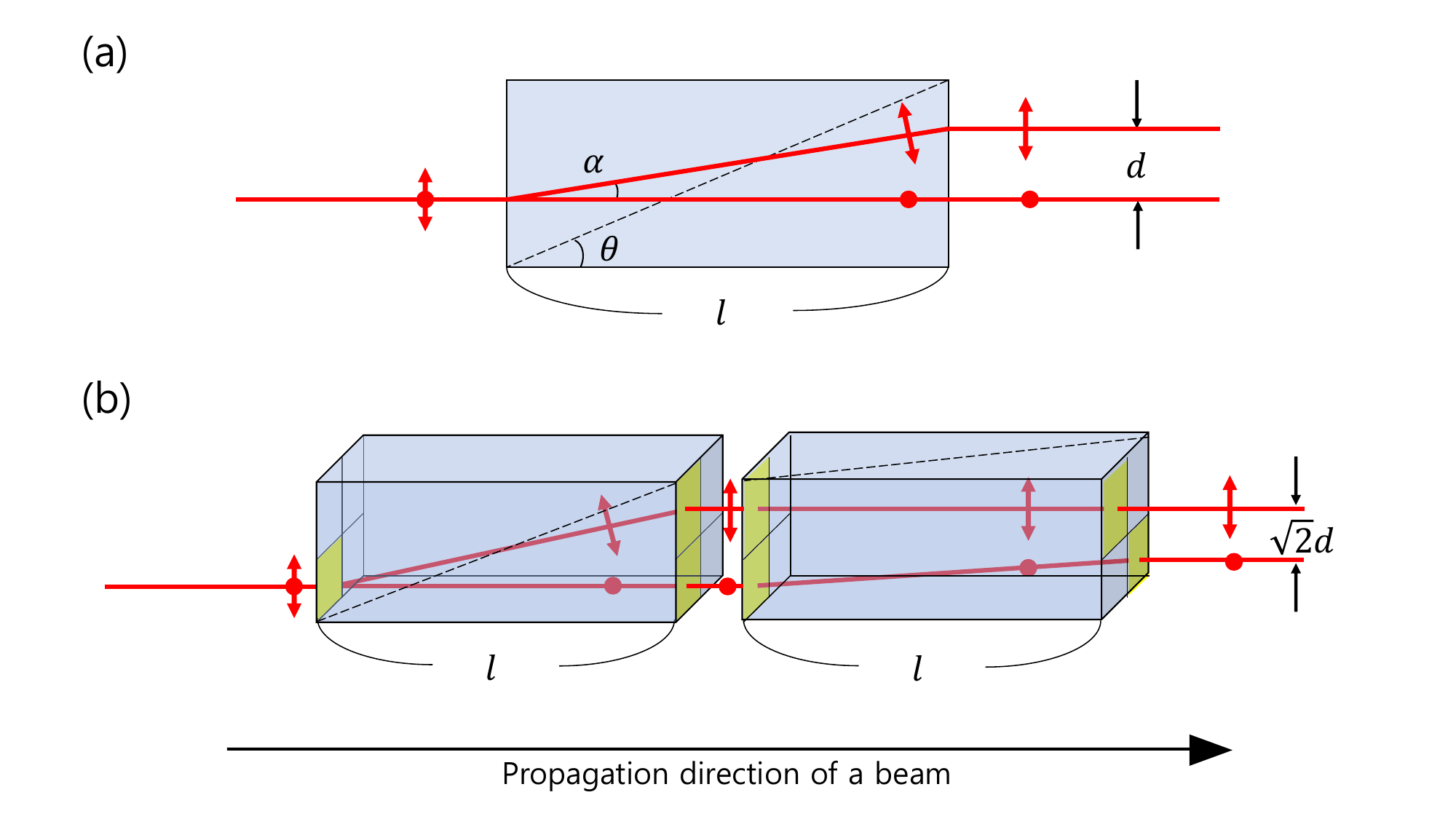}
\caption{Beam displacer action in longitudinal-transverse axes. The dashed line corresponds to the beam displacer's optical axis. (a) Action of one beam displacer, and (b) action of two beam displacers with their optical axes rotated by 90 degrees.}
\label{Fig:BD-explanation}
\end{figure}

On the one hand, to generate polarization-entanglement using interferometers, one might require dual-wavelength optical components, active stabilization, and demanding alignment. An alternative configuration that relaxes these requirements was introduced by M. Fiorentino and R. G. Beausoleil~\cite{fiorentino2008compact} by employing beam displacers (BD). These devices are polarization-dependent birefringent materials that produce an angular deviation on an impinging optical field, see Fig.~\ref{Fig:BD-explanation}. Therefore, after the beam has traversed a BD, it can be spatially separated into two transverse locations, where the distance separation depends on the field polarization, its wavelength, and the BD's crystal length and material. Polarization-entanglement sources using beam displacers have been experimentally tested for degenerated~\cite{evans2010bright} and non-degenerated SPDC~\cite{horn2019auto} and for non-degenerated four-wave mixing~\cite{lee2021sagnac}. On the other hand, spatial entanglement~\cite{walborn2010spatial} is generated in bulk crystals, e.g. beta barium
borate (BBO) or periodically poled potassium titanyl phosphate (PPKTP). The quality of this entanglement is enhanced by reducing the crystals' length, where the shortest available crystals in the market are in the order of millimeters. These crystals can fulfill the regime of \textit{thin crystals} introduced by C. H. Monken \textit{et al.}~\cite{monken1998trasnfer}, where the bi-photon amplitude is proportional to the angular spectrum of the pump and, thus, the pump beam waist determines the quality of the bi-photon spatial correlations. Our approach utilizes double displacements~\cite{horn2019auto} of non-degenerated photons and uses two thin PPKTP crystals for SPDC. Due to these properties, our source is ideal for quantum imaging applications where polarization plays an important role in the illumination.
To form an imaging plane, either far field or near field, one needs to employ an optical system from the source where the photon pairs are being emitted. If the emission cones from two different sources are not in the same transverse plane, like the case of different approaches to polarization entangled photon-pair sources, e.g. the cross-crystal configuration, it is not possible to form a unique imaging plane that corresponds to the superimposition of both independent imaging planes. With our double BD approach, since both crystals are independent, their longitudinal positions can be tuned to ensure spatial indistinguishability.


\begin{figure}[htpb]
    \centering
    \includegraphics[width=8cm]{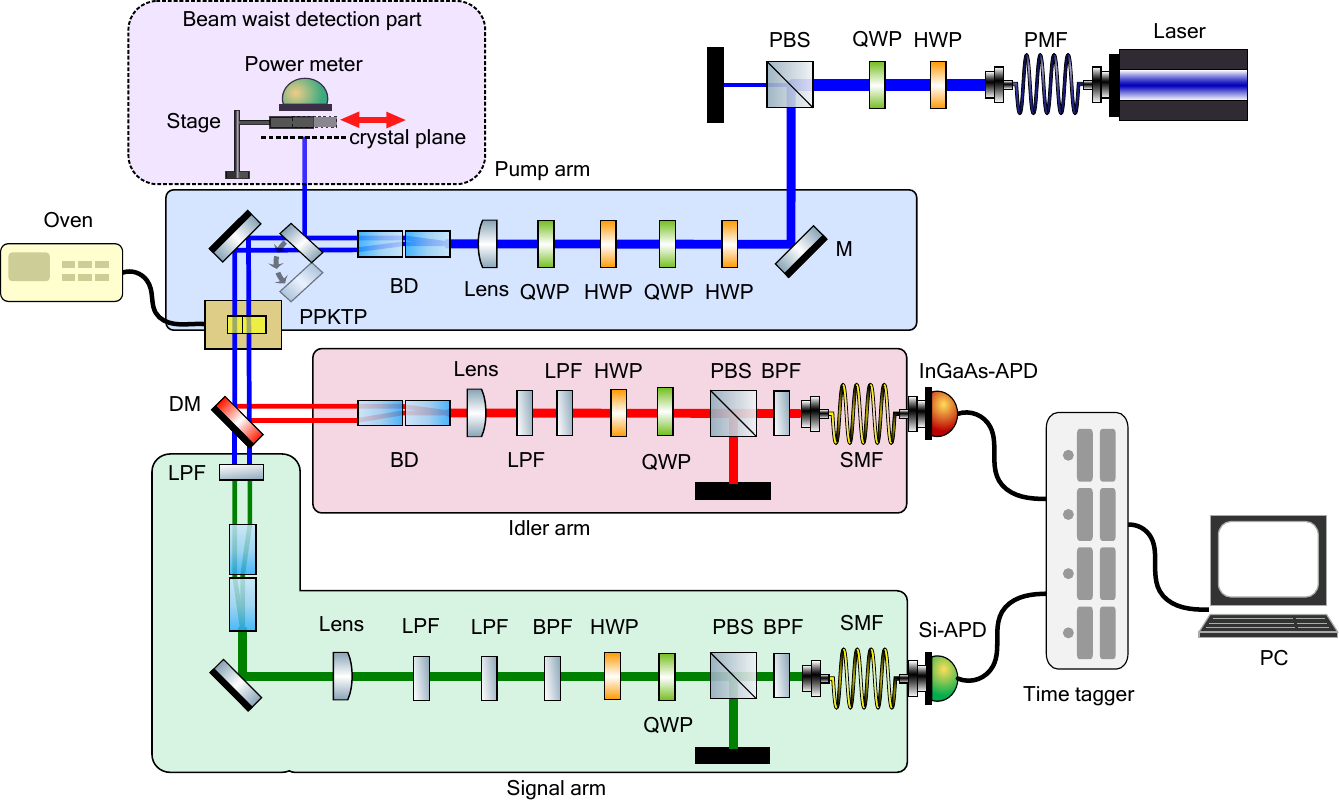}
    \caption{Experimental setup. Signal and idler photon pairs are generated via type-0 SPDC at $\lambda_s=548.2$ nm and $\lambda_i=1550$ nm, respectively. 
    Two BDs recombine the signal and idler optical paths and, therefore, erase their path information resulting in the state in~\eqref{Eq:Phi_plus}. Both photons traverses a polarization analysis module. Coincidences of signal-idler photons are counted with a time-tagging module. More details in the main text.}
    \label{Fig:Setup}
\end{figure}

In terms of polarization the desired state shall be a Bell state~\cite{einstein1935EPR,bell1964theorem} of the form 
\begin{equation}
    \ket{\Phi^-}=\frac{1}{\sqrt{2}}\left(\ket{H_s}\ket{H_i}-\ket{V_s}\ket{V_i}\right).
    \label{Eq:Phi_plus}
\end{equation}
Here, s (i) stands for signal (idler) photon, and $H$ and $V$ are horizontal and vertical polarizations, respectively. Our experimental setup is depicted in Fig.~\ref{Fig:Setup}. For generating the state in ~\eqref{Eq:Phi_plus}, we start with a CW 405 nm pump laser that traverses a combination of plate retarders in order to change its polarization and the relative phase between $H$ and $V$. We initially prepare the pump beam in the diagonal polarization ($D$). Later, a lens focused the pump beam at the source's plane. A flip mirror placed before the nonlinear crystal, allows to alternatively send the beam to a beam waist measurement module, consisting of a 1D translation stage, placed exactly at the same position of the crystal plane in the main setup, and a power meter. Before reaching the nonlinear crystals, the two components of the pump beam, $H$ and $V$, are spatially separated by two consecutive BDs. The BD has two optical axes: the ordinary axis (polarization independent with refractive index --- $n_o$) and the extra-ordinary axis (polarization dependent with refractive index--- $n_e$). Thus, the impinging optical field is split into two rays inside the BD, forming an angle $\alpha$; see Fig.~\ref{Fig:BD-explanation}(a). This angular deviation depends on the polarization and the BD's angle $\theta$ by 
\begin{align}
    \tan(\alpha) = \left(  \frac{1- n^2_o}{n^2_e} \right) \, \frac{\tan(\theta)}{(1 + n^2_o/n^2_e)  \tan^2(\theta)},
    \label{Eq:angle-deviation}
\end{align}
where $(n^2_e (\theta))^{-1}= \cos^2(\theta)/n^2_o+ \sin^2(\theta)/n^2_e$. In our experiment, the axes of the two consecutive BDs are rotated by 90 degrees~\cite{horn2019auto}. By this, the two polarizations ($H$ and $V$) are displaced into two orthogonal directions (e.g., `up' and `left') in the transverse plane, and in this way, no extra phase needs to be compensated; see Fig.~\ref{Fig:BD-explanation}(b). In our experiment, we aim for a beam separation of $d=0.5$ mm, thus obtaining a total separation of $\sim 0.71$ mm after crossing the two BDs. The beam displacers' lengths can be inferred from the Sellmeier equation that relates the refractive indices and wavelengths of the fields involved by 
\begin{align}
    n^2(\lambda)=K + \frac{Q_1\, \lambda^2}{\lambda^2-P_1}+\frac{Q_2\, \lambda^2}{\lambda^2-P_2}.
    \label{Eq:Sellmeier}
\end{align}
For alpha-barium borate, 
the parameters in ~\eqref{Eq:Sellmeier} correspond to $K=2.67579$, $Q_1= 0.02009$, $Q_2=0.00528$, $P_1=0.00470$, and $P_2=0$, which are quoted for the wavelength in micrometers. Using ~\eqref{Eq:angle-deviation} and~\eqref{Eq:Sellmeier}, and $\lambda_p=405$ nm, we obtained the crystal length $l_p=$12.52 mm for the pump beam.
After the action of the first two BDs, the pump beam has been split into two spatially separated positions, where each of which carries an orthogonal polarization with respect to the other. These pump beams illuminate two nonlinear crystals to produce photon pairs through a type-0, non-degenerated SPDC process. The pump ($\lambda_s=405$ nm) photon is down-converted to a signal ($\lambda_s=548.2$ nm) and an idler ($\lambda_i=1550$ nm) photons. Both sources are identical; PPKTP crystal with dimensions of 1$\times$2$\times$2 mm$^3$ (width$\times$height$\times$length) and a poling period of 4.225 $\mu$m. The crystals are in a cross-crystal configuration (in the transverse plane) such that one of them produces the states $\ket{H_s}\ket{H_i}$ while the other generates $\ket{V_s}\ket{V_i}$. Signal and idler photons are split with the help of a dichroic mirror. The reflected idler beams are recombined with two identical BDs of length $l_i$=13.66 mm. Similarly, the transmitted signal beams are recombined with two identical BDs of length $l_s$=12.98 mm. In this way, the path information of the signal-idler pair is erased, and the Bell state in ~\eqref{Eq:Phi_plus} is generated.

In the polarization analysis modules, there is a cascade of a half-wave plate (HWP$_{s/i}$), a quarter-wave plate (QWP$_{s/i}$), and a polarizing beam splitter (PBS$_{s/i}$) in signal and idler arms. Since QWPs are employed only for tomography reconstruction, we will assume they are set at their fast axis until then. Each PBS is oriented in such a way that it transmits horizontal and reflects vertical polarizations. Signal (idler) photon is spectrally cleaned with an interference filter with a band-pass of 548$\pm$6 nm (1550$\pm$3 nm) before being collected by a single-mode fiber placed at the horizontal output of its PBS. Signal photons are detected with a single-photon avalanche diode 
device with a detection efficiency of 55$\%$. For idler photons, an indium gallium arsenide (InGaAs) device with a detection efficiency of 25$\%$ is used. Coincidence count rates ($C(s,i)\propto \bra{\Phi^-} E^-_i  E^-_s  E^+_i E^+_s \ket{\Phi^-}$, E: electric field operator) of signal-idler pairs are counted with a time-tagging module. 

\begin{figure}[htpb]
    \centering
    \includegraphics[width=8cm]{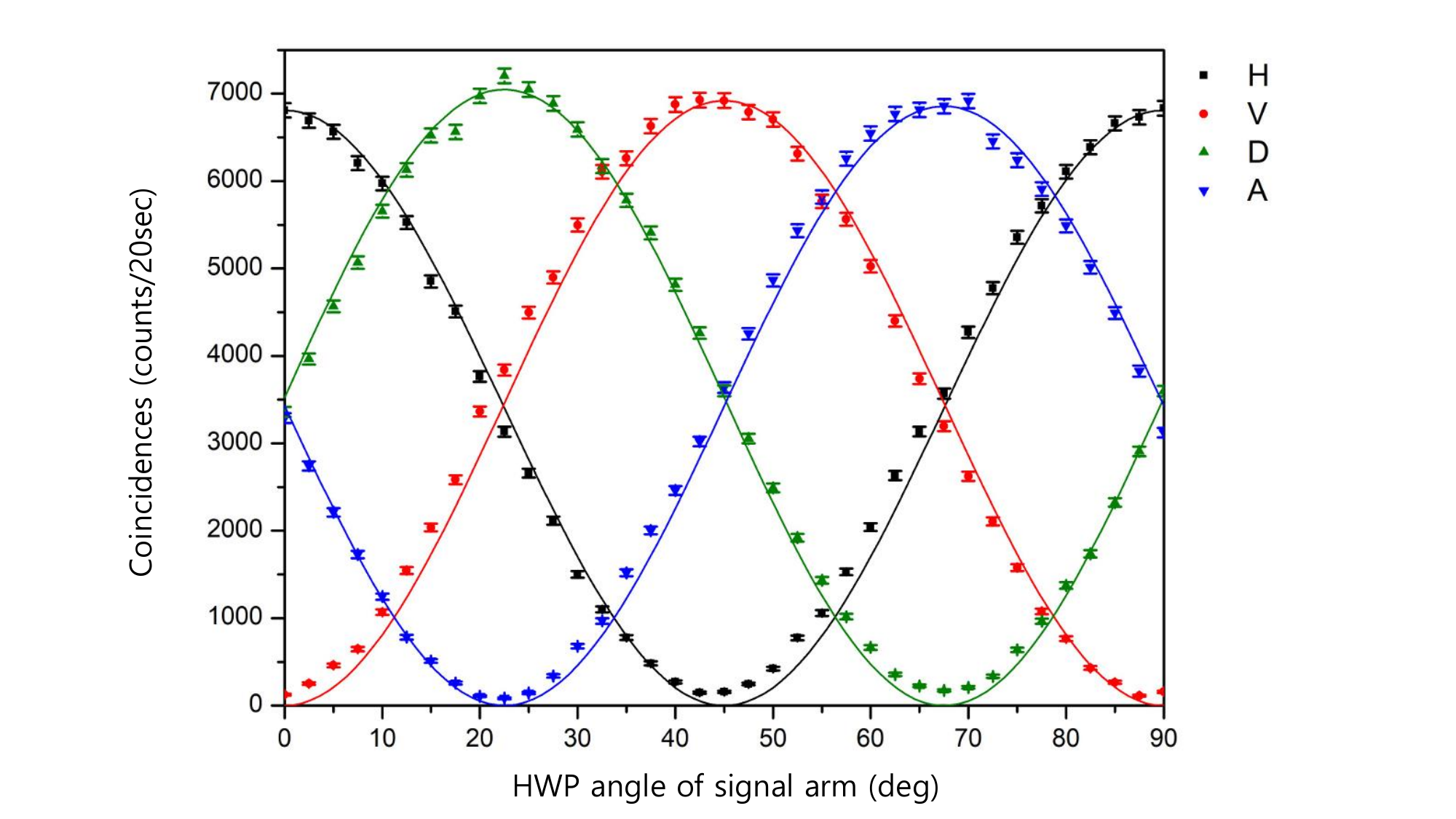}
    \caption{Coincidence detection in each basis. Dots are experimental data and lines are theoretical fitting. Colors differ by the polarization of idler photon polarization: black - horizontal polarization, red - vertical polarization, green - diagonal polarization, blue - anti-diagonal polarization.}
    \label{fig:vis-results}
\end{figure}

To verify polarization entanglement of the bi-photon state, we performed different measurements that are prepared by changing the angle of the retarders HWP$_s$ and HWP$_i$ and checking for the coincidences. The first measurement corresponds to the visibility ($\mathcal{V}$) between the signal-idler pair in the $H/V$ and $D/A$ bases, where $A$ stands for anti-diagonal polarization. On $H/V$ basis, we have \cite{ortega2021multicore}
\begin{align}
    \mathcal{V}_{H/V}= \frac{C(H,H)+C(V,V)-C(H,V)-C(V,H)}{C(H,H)+C(V,V)+C(H,V)+C(V,H)},
\end{align}
where $C(m,n)$ corresponds to a signal-idler coincidence detection in the transmitted outputs of the PBSs at HWPs' angles $m$ and $n$ in signal and idler paths, respectively. These coincidence values are obtained by placing the retarders HWP$_s$ and HWP$_i$ at angles ($H=0^\circ$, $V=45^\circ$). The visibility in the D/A basis is similarly computed; however, the angles of HWP$_s$ and HWP$_i$ are changed to ($D=22.5^\circ$, $A=-22.5^\circ$). In $H/V$ basis we obtained a visibility of $\mathcal{V}_{H/V}=98.4\pm0.1$ whereas in $D/A$ basis $\mathcal{V}_{D/A}=94.7\pm0.2$. Figure~\ref{fig:vis-results} depicts the experimental coincidences plot for a fixed HWP$_i$ angle while HWP$_s$ is varied. 

In a second experiment, we test the Clauser-Horne-Shimony-Holt (CHSH) inequality~\cite{clauser1969proposed}, which is a standard technique to quantify entanglement in a bipartite quantum system. The CHSH inequality can be expressed as 
\begin{equation}
    \MS= \E (m_1,n_1)+\E (m_2,n_2)+\E (m_1,n_2)-\E (m_2,n_1),    
\end{equation}
where
\begin{equation}
 \E (m,n) =   \frac{C(m,n)+C(m^{\perp},n^{\perp})-C(m^{\perp},n)-C(m,n^{\perp})}{C(m,n)+C(m^{\perp},n^{\perp})+C(m^{\perp},n)+C(m,n^{\perp})}.
\end{equation}
 $C(m,n)$ still corresponds to a joint detection between the transmitted outputs of the PBSs at HWPs' angles $m$ and $n$, while $m^{\perp}$ and $n^{\perp}$ denote the reflected outputs of the PBSs for the same angles. Since we used only the transmitted output of the PBSs, the angles employed in our setup for the CHSH inequality are $(m_1=0^\circ, m^{\perp}_1=45^\circ,m_2=22.5^\circ,m^{\perp}_2=67.5^\circ)$ for HWP$_s$, and $(n_1= 11.25^\circ, n^{\perp}_1=33.75^\circ, n_2=56.25^\circ, n^{\perp}_2=78.25^\circ)$ for  HWP$_i$. By doing this, we obtained a value for the CHSH inequality of $\MS=2.75\pm0.01$, demonstrating a high violation from the classical limit of $\MS=2$ by 75 standard deviations.
In a last experiment regarding polarization entanglement, we perform a quantum state tomography~\cite{james2001measurement} of the Bell state in ~\eqref{Eq:Phi_plus}. To reconstruct the density matrix, and following previous notation, we measure signal-idler coincidences using the following combinations $C(H,H)$, $C(H,D)$, $C(H,V)$, $C(H,L)$, $C(V,H)$, $C(V,D)$, $C(V,V)$, $C(V,L)$, $C(D,H)$, $C(D,D)$, $C(D,V)$, $C(D,R)$, $C(R,H)$, $C(R,D)$, $C(R,V)$, and $C(R,L)$. $R$ ($L$) stands for right (left) circular polarized light and can be prepared by rotating a QWP at $45^\circ$ ($-45^\circ$). In this way, the reconstructed density matrix $\hat{\rho}_{\textup{exp}}$ is 
\begin{figure}[htbp]
    \centering
\includegraphics[width=9cm]{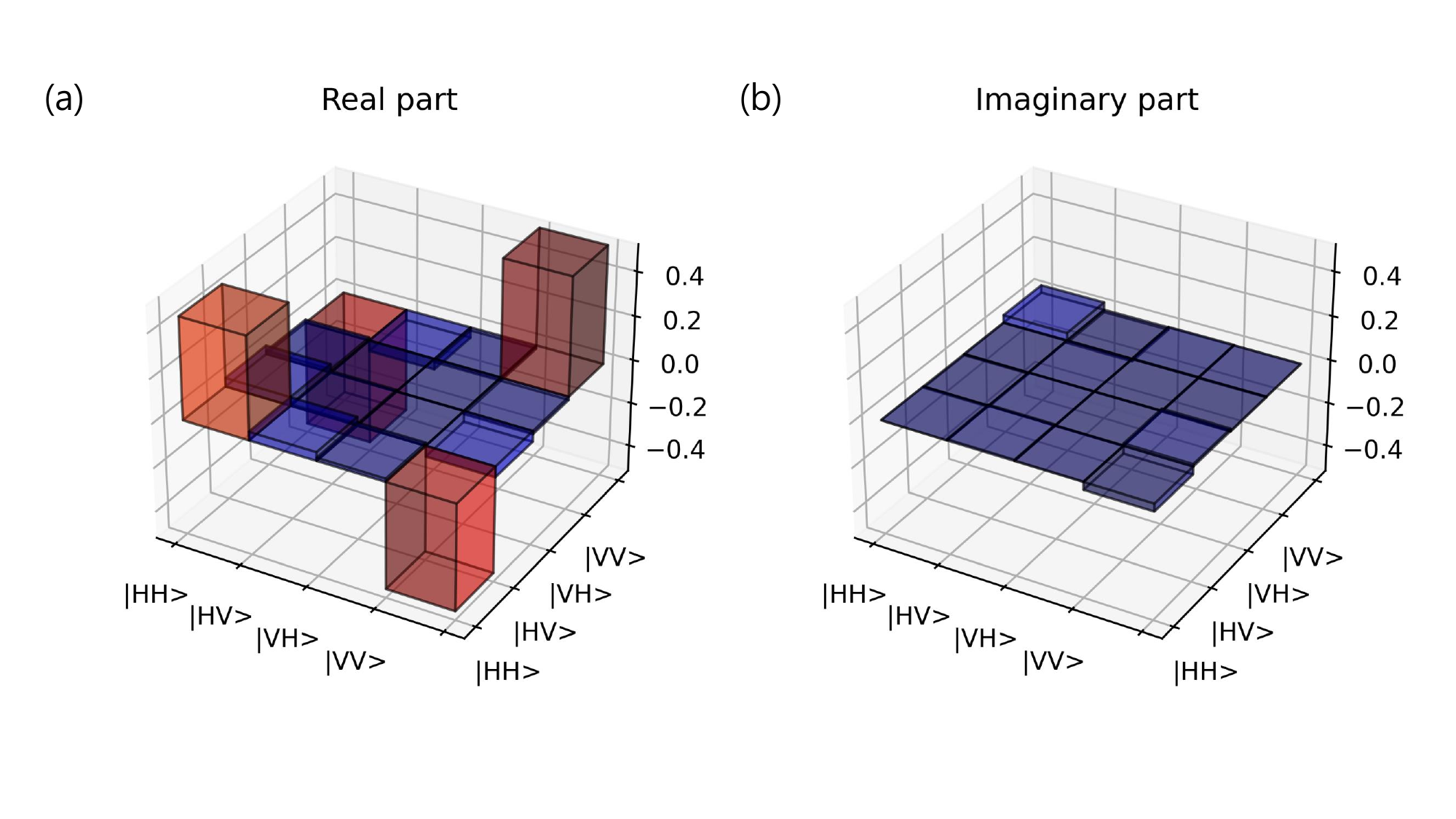}
    \vspace{-3.5em}
    \caption{Density matrix reconstruction. (a) Real part and (b) imaginary part.}
    \label{fig:real}
\end{figure}

{
\tiny
\begin{align}
\hat{\rho}_{\textup{exp}}=
    \begin{pmatrix}
      0.4580 & 0.0386+i0.0050 & 0.0142+i0.0014 & -0.4802+i0.0348 \\
       0.039-i0.0050 & 0.0078 & 0.0002-i0.0005 & -0.048+i0.0107 \\
       0.0014-i0.0014 & 0.0002+i0.0005 & 0.0007 & -0.0144-i0.0022 \\
      -0.4802-i0.0348 & 0.0481-i0.01069 & -0.01445+i0.0022 & 0.5335
    \end{pmatrix}.
\label{rho}
\end{align}
}

The fidelity of our state, $F(\hat{\rho}_{\textup{exp}},\hat{\rho}_{\textup{th}})=\left\{\textrm{tr}\left(\sqrt{\sqrt{\hat{\rho}_{\textup{exp}}}\hat{\rho}_\textup{th}\sqrt{\hat{\rho}_{\textup{exp}}}} \right) \right\}^2$, where $\hat{\rho}_\textup{th}= \ket{\Phi^-} \bra{\Phi^-}$ is the density matrix of ~\eqref{Eq:Phi_plus}, is $F=0.975 \pm 0.004$. The error was obtained by the Gaussian estimation. Figures~\ref{fig:real}(a) and~~\ref{fig:real}(b) show the real part and the imaginary part of $\hat{\rho}_{\textup{exp}}$, respectively. Table \ref{table1} presents a comparison of our source’s polarization characteristics along with other sources using BDs. Our source shows overall good visibilities and polarization entanglement when compared with these sources.  In the following, we study the spatial properties of our source.

\begin{table}[]
\caption{Comparison between different polarization entangled photon-pair sources using beam displacers.\label{table1}}
\resizebox{\columnwidth}{!}{%
\begin{tabular}{|
>{\columncolor[HTML]{EFEFEF}}c |c|c|c|c|
>{\columncolor[HTML]{EFEFEF}}c |}
\hline
References       & Fiorentino\cite{fiorentino2008compact}  & Horn\cite{horn2019auto}       & Lohrmann\cite{lohrmann2020broadband}     & Sziachetka\cite{szlachetka2023ultrabright}     & This source  \\ \hline
Configuration    & \begin{tabular}[c]{@{}c@{}}type-II\\ non-degenerate\\ single BD\end{tabular} & \begin{tabular}[c]{@{}c@{}}type-0\\ degenerate\\ double BDs\end{tabular} & \begin{tabular}[c]{@{}c@{}}type-0\\ degenerate\\ single BD\end{tabular} & \begin{tabular}[c]{@{}c@{}}type-0\\ non-degenerate\\ double BDs\end{tabular} & \begin{tabular}[c]{@{}c@{}}type-0\\ non-degenerate\\ double BDs\end{tabular} \\ \hline
Crystal          & PPKTP                                                                        & PPLN                                                                     & PPKTP                                                                   & MgO:PPLN                                                                     & PPKTP                                                                        \\ \hline
Crystal length   & 10mm                                                                         & 30mm                                                                     & 10mm                                                                    & 20mm                                                                         & 2mm                                                                          \\ \hline
Visibility (H/V) & (98$\pm$4)\%                                                                 & -                                                                        & (99.0$\pm$0.2)\%                                                        & (96.49$\pm$0.76)\%                                                           & (98.4$\pm$0.1)\%                                                                 \\ \hline
Visibility (D/A) & (81$\pm$1)\%                                                                 & -                                                                        & (96.4$\pm$0.4)\%                                                        & (96.01$\pm$0.81)\%                                                           &(94.7$\pm$0.2)\%                                                                \\ \hline
S parameter      & 2.54$\pm$0.01                                                                & -                                                                        & -                                                                       & 2.71$\pm$0.06                                                                & 2.75$\pm$0.01                                                                \\ \hline
Fidelity         & -                                                                            & 98\%                                                                     & -                                                                       & (96.72$\pm$0.01)\%                                                           & (97.5$\pm$0.4)\%                                                           \\ \hline
\end{tabular}%
}
\end{table}

In quantum imaging, there are plenty of techniques to quantify or estimate the bi-photon spatial correlations~\cite{law2004analysis,lorenzo2009direct,howell2004epr,achatz2022certifying,just2013transverse,edgar2012imaging}. Here, we decide to include a parameter that is associated with the source's properties. The number of spatial modes produced by a crystal of length $l_c$ and refractive index $n_s$ ($n_i$) for signal (idler) photon can be estimated by~\cite{kviatkovsky2020microscopy}
\begin{align}
    N_{\textup{modes}}=\frac{5.56 \, \pi \, w_p^2 \, n_i \, n_s}{ \ln(2) \, l_c\, (\lambda_i n_s+ \lambda_s n_i)}
    \label{Eq:mode}
\end{align}
$w_p$ stands for the pump waist. Evidently, $N$ in ~\eqref{Eq:mode}, can be enhanced by decreasing the length of the nonlinear crystals. We perform an estimation of the spatial modes produced by our specific source. For this, we measured $w_p$  with a transverse translation stage and a power meter.
\begin{figure}[ht]
    \centering
    \includegraphics[width=8cm]{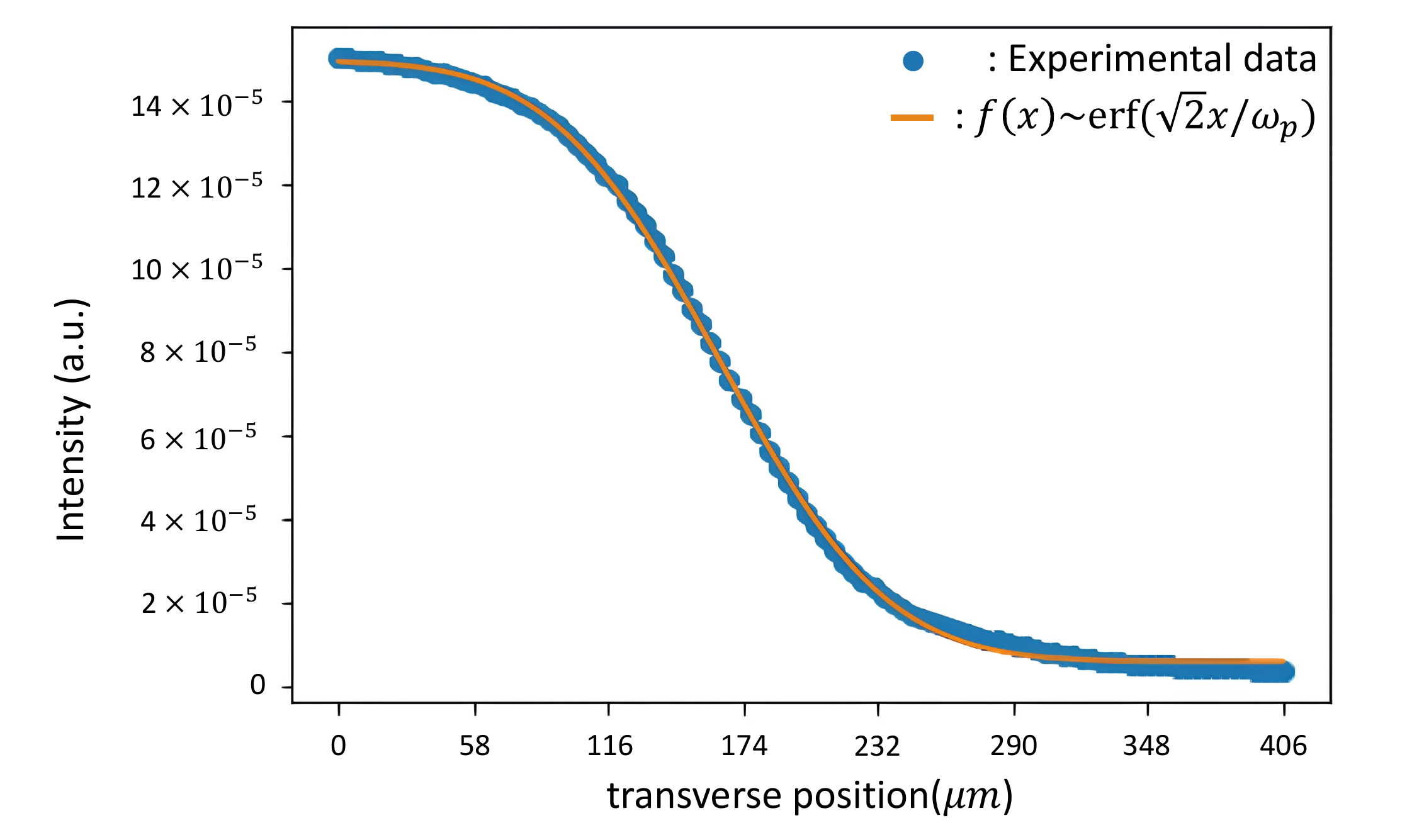}
    \caption{Measurement of the beam waist by a translation stage and a power meter. Intensity was measured at each transverse position (blue dots). The beam waist was estimated by fitting an Error function (orange line). }
    \label{fig:beam_waist}
\end{figure}
Figure~\ref{fig:beam_waist} shows the experimental results for the beam waist measurement, where we obtained $w_p=229\pm2.5$ $\mu$m from an error function fitting; see inset. Thus, the number of spatial modes estimated by ~\eqref{Eq:mode} was $550\pm12$. It is noteworthy that our source implies a large number of modes. Sharper spatial correlations increase with the number of spatial modes, meaning that our source has promising applications in quantum imaging. The exploitation of a high number of independent modes in the same frame from a quantum-light source is the paramount requirement for quantum enhanced wide field imaging. More specifically, the number of features of a sample that can be inferred within a single frame, is determined by the number of spatial modes enclosed in the illuminating field \cite{samantaray2017realization}. A larger number of spatial modes can be achievable by increasing the pump waist size, which is limited by the crystal aperture of $1\times2$ mm$^2$ in our case.

We have experimentally introduced a new polarization-entangled photon pair source using beam displacers for quantum imaging applications. The novelty of our experiment resides in the wavelengths employed ( $\lambda_s=548.2$ nm and $\lambda_i=1550$ nm), the tunable spatial indistinguishability for wide-field imaging, and the enhancement of spatial modes employing thin nonlinear crystals in comparison to other sources using BDs, see Table \ref{table1}.  
In our source, the spatial correlations (as well as the number of spatial modes) can be improved by (1) increasing the aperture 
or (2) reducing the crystal length \cite{fuenzalida2024nonlinear}. For the former, BBO crystals with large apertures are already available on the market. Additionally, for higher photon pair generation, PPKTP crystals with apertures of up to $4\times4$ mm$^2$ have recently been introduced~\cite{lee2023large}. For the latter, nonlinear metasurfaces can be built in lengths in the order of microns~\cite{marino2019spontaneous}, which might be an option for reaching the diffraction limit~\cite{vega2022fundamental} in techniques such as ghost imaging~\cite{pittman1995optical} and quantum imaging with undetected photons~\cite{lemosquantum2014}. Of particular importance, our source can be useful for imaging objects with birefringent properties~\cite{brasselet2023polarization} or quantum imaging distillation~\cite{fuenzalida2023experimental}. In other areas, such as quantum communication, it has been proven that distillation can be enhanced by exploiting hyper-entangled states. Hence, our developed source can offer more resilience to noise than techniques only using spatial correlations. Lastly, we expect to perform a thorough characterization of the spatial properties and, thus, corroborate hyper-entanglement in polarization and space in future experiments.

\begin{backmatter}
\bmsection{Funding} This research was carried out within the scope of the ICON-Programme of the Fraunhofer Society „Integrated Photonic Solutions for Quantum Technologies (InteQuant)”. We also acknowledge support from the European Union’s Horizon 2020 Research and Innovation Action under Grant Agreement No. 101113901 (Qu-Test, HORIZON-CL4-2022-QUANTUM-05-SGA).  

\bmsection{Disclosures} The authors declare no conflicts of interest.

\bmsection{Data Availability Statement} Data underlying the results presented in this paper are not publicly available at this time but may be obtained from the authors upon
reasonable request.

\end{backmatter}

\bibliography{sample}



\ifthenelse{\equal{\journalref}{aop}}{%
\section*{Author Biographies}
\begingroup
\setlength\intextsep{0pt}
\begin{minipage}[t][6.3cm][t]{1.0\textwidth} 
  \begin{wrapfigure}{L}{0.25\textwidth}
    \includegraphics[width=0.25\textwidth]{john_smith.eps}
  \end{wrapfigure}
  \noindent
  {\bfseries John Smith} received his BSc (Mathematics) in 2000 from The University of Maryland. His research interests include lasers and optics.
\end{minipage}
\begin{minipage}{1.0\textwidth}
  \begin{wrapfigure}{L}{0.25\textwidth}
    \includegraphics[width=0.25\textwidth]{alice_smith.eps}
  \end{wrapfigure}
  \noindent
  {\bfseries Alice Smith} also received her BSc (Mathematics) in 2000 from The University of Maryland. Her research interests also include lasers and optics.
\end{minipage}
\endgroup
}{}

\end{document}